\documentclass[aps,prb,twocolumn,showpacs,superscriptaddress]{revtex4}
\usepackage{graphicx}
\usepackage{bm}
\usepackage{color}

\begin{document}

\title{The effect of non-superconducting dopants (Mn, V, Cr and Cu) on the nematic fluctuations in iron-based superconductors}
\author{Yanhong Gu}
\affiliation{Beijing National Laboratory for Condensed Matter Physics, Institute of Physics, Chinese Academy of Sciences, Beijing 100190, China}
\affiliation{School of Physical Sciences, University of Chinese Academy of Sciences, Beijing 100190, China}
\author{Yuan Wei}
\affiliation{Beijing National Laboratory for Condensed Matter Physics, Institute of Physics, Chinese Academy of Sciences, Beijing 100190, China}
\affiliation{School of Physical Sciences, University of Chinese Academy of Sciences, Beijing 100190, China}
\author{Dongliang Gong}
\affiliation{Beijing National Laboratory for Condensed Matter Physics, Institute of Physics, Chinese Academy of Sciences, Beijing 100190, China}
\affiliation{School of Physical Sciences, University of Chinese Academy of Sciences, Beijing 100190, China}
\author{Wenliang Zhang}
\affiliation{Beijing National Laboratory for Condensed Matter Physics, Institute of Physics, Chinese Academy of Sciences, Beijing 100190, China}
\affiliation{School of Physical Sciences, University of Chinese Academy of Sciences, Beijing 100190, China}
\author{Wenshan Hong}
\affiliation{Beijing National Laboratory for Condensed Matter Physics, Institute of Physics, Chinese Academy of Sciences, Beijing 100190, China}
\affiliation{School of Physical Sciences, University of Chinese Academy of Sciences, Beijing 100190, China}
\author{Xiaoyan Ma}
\affiliation{Beijing National Laboratory for Condensed Matter Physics, Institute of Physics, Chinese Academy of Sciences, Beijing 100190, China}
\affiliation{School of Physical Sciences, University of Chinese Academy of Sciences, Beijing 100190, China}
\author{Xingguang Li}
\affiliation{Department of Physics and Beijing Key Laboratory of Opto-electronic Functional Materials \& Micro-nano Devices, Renmin University of China, Beijing 100872, P. R. China}
\author{Congkuan Tian}
\affiliation{Department of Physics and Beijing Key Laboratory of Opto-electronic Functional Materials \& Micro-nano Devices, Renmin University of China, Beijing 100872, P. R. China}
\author{Peng Cheng}
\affiliation{Department of Physics and Beijing Key Laboratory of Opto-electronic Functional Materials \& Micro-nano Devices, Renmin University of China, Beijing 100872, P. R. China}
\author{Hongxia Zhang}
\affiliation{Department of Physics and Beijing Key Laboratory of Opto-electronic Functional Materials \& Micro-nano Devices, Renmin University of China, Beijing 100872, P. R. China}
\author{Wei Bao}
\affiliation{Department of Physics and Beijing Key Laboratory of Opto-electronic Functional Materials \& Micro-nano Devices, Renmin University of China, Beijing 100872, P. R. China}
\author{Guochu Deng}
\affiliation{Australian Centre for Neutron Scattering, Australian Nuclear Science and Technology Organisation, New Illawarra Road, Lucas Heights, NSW 2234, Australia}
\author{Xin Li}
\affiliation{Key Laboratory of Neutron Physics and Institute of Nuclear Physics and Chemistry, China Academy of Engineering Physics, Mianyang 621999, China}
\author{Jianming Song}
\affiliation{Key Laboratory of Neutron Physics and Institute of Nuclear Physics and Chemistry, China Academy of Engineering Physics, Mianyang 621999, China}
\author{Yi-feng Yang}
\affiliation{Beijing National Laboratory for Condensed Matter Physics, Institute of Physics, Chinese Academy of Sciences, Beijing 100190, China}
\affiliation{School of Physical Sciences, University of Chinese Academy of Sciences, Beijing 100190, China}
\affiliation{Songshan Lake Materials Laboratory , Dongguan, Guangdong 523808, China}
\author{Huiqian Luo}
\affiliation{Beijing National Laboratory for Condensed Matter Physics, Institute of Physics, Chinese Academy of Sciences, Beijing 100190, China}
\affiliation{Songshan Lake Materials Laboratory , Dongguan, Guangdong 523808, China}
\author{Shiliang Li}
\email{slli@iphy.ac.cn}
\affiliation{Beijing National Laboratory for Condensed Matter Physics, Institute of Physics, Chinese Academy of Sciences, Beijing 100190, China}
\affiliation{School of Physical Sciences, University of Chinese Academy of Sciences, Beijing 100190, China}
\affiliation{Songshan Lake Materials Laboratory , Dongguan, Guangdong 523808, China}
\begin{abstract}
We have systematically studied the nematic susceptibility in non-superconducting Ba(Fe$_{1-x}$TM$_{x}$)$_2$As$_2$ (TM = Cr, Mn, V and Cu) by measuring the uniaxial pressure dependence of the resistivity along the Fe-As-Fe direction. The nematic susceptibilities in all samples show the Curie-Weiss-like behavior at high temperatures, where the nematic Curie constant $A_n$ can be derived, similar to the Curie constant in a paramagnetism. While all these dopants do not introduce superconductivity in BaFe$_2$As$_2$, their effects on nematic fluctuations are different. In Mn, Cr and V doped samples, $|A_n|$ decreases significantly with the increasing doping level. On the other hand, $|A_n|$ increases dramatically with Cu doping, similar to the superconducting Ni-doped BaFe$_2$As$_2$. However, the nematic susceptibility is suppressed at low temperatures for $x$ larger than $0.04$, which may be related to the short-range antiferromagnetic order that survives up to very high doping level. Doping Mn, Cr and Cu into the optimally-doped superconducting BaFe$_2$(As$_{0.69}$P$_{0.31}$)$_2$ also strongly reduces $|A_n|$. Compared with those systems that clearly exhibit superconductivity, such as Ni, K or P doped samples, our results suggest a strong connection between the nematic and spin degrees of freedom. Moreover, the reason of the suppression of superconductivity by dopants such as Cr, Mn, V and Cu may be correlated with the suppression of nematic fluctuations.  
\end{abstract}



\maketitle

\section{introduction}
Iron-based superconductors have attracted many interests due to their high-temperature superconductivity, but the underlying mechanism is still unclear \cite{SiQ16,DaiP15,ScalapinoDJ12}. Compared to cuprates and many other unconventional superconductors, a particular interesting fact is that iron-based superconductivity can be achieved by various dopants at different sites \cite{HosonoH15}. For example, BaFe$_2$As$_2$ (Ba-122) shows both the antiferromagnetic (AFM) and tetragonal-to-orthorhombic structural transitions with a collinear or stripe magnetic structure \cite{HuangQ08}. Taking it as the parent compound, superconductivity can be achieved by substituting Fe by Ni, Co, Ru and Rh, or Ba by K and Na, or As by P, and so on \cite{SefatAS08,LiLJ09,NiN09,SharmaS10,RotterM08,CortesGilR10,JiangS09,LuoH12}. However, one cannot always obtain superconductivity by the substitution, as shown in Mn, Cr, V and Cu doped systems \cite{NiN10,ThalerA11,LiX18}. In both Mn and Cr doped Ba-122, the stripe AFM structure changes to G-type structure with the competition between stripe-type and G-type spin fluctuations \cite{KimMG10,MartyK11,TuckerGS12,InosovDS13}, most likely because the magnetic structures in both BaMn$_2$As$_2$ and BaCr$_2$As$_2$ are G-type \cite{YogeshS09,FilsingerKA17}. In V and Cu doped Ba-122, the AFM orders are gradually suppressed to short-range or spin-glass-like state without the appearance of superconductivity \cite{KimMG12,KimMG15,TakedaH14,WangW17,LiX18}. It is not only that doping these dopants cannot lead to superconductivity but also that they can quickly suppress superconductivity in superconducting samples \cite{NiN09,ChengP10,LiJ12,ZhangR14,ZhangW19}. It has been suggested that the above results are due to their local effects on the electronic and magnetic properties \cite{ChengP10,TuckerGS12,OnariS09,TakedaH14,TexierY12,SuzukiH13,IdetaS13,WangX14,UrataT15,KobayashiT16,GastiasoroMN14}. Since these explanations depend on the detailed effects of particular dopants, there is a lack of comprehensive understanding on the physics that is directly associated with superconductivity. This is not surprising because we have not found a general picture for the superconductivity itself yet. 

Dopants affect not only antiferromagnetism and superconductivity, but also nematicity. The nematic order in iron-based superconductors breaks the rotational symmetry of the tetragonal lattice and is believed to result in the tetragonal-to-orthorhombic structural transition at $T_s$ \cite{FernandesRM14}. While its mechanism is still under debates \cite{FernandesRM14}, the nematic order and its fluctuations can be observed by the anisotropic properties revealed by many different techniques \cite{ChuJH10,ChuJH12,YiM11,NakajimaM11,LuX14,BohmerAE14,ZhangW16,LiuZ16}. There are increasing evidences both theoretically and experimentally that nematic fluctuations may be important for superconductivity \cite{FernandesRM13,LedererS15,KleinA18,kushnirenkoYS18,KuoHH16,GuY17}. We have shown in a previous study that the enhancement of nematic fluctuations seems to be associated with both the suppression of the AFM ordered moment and the appearance of superconductivity \cite{GuY17}. These observations are made in superconducting systems, here we further study the correlation between nematic fluctuations and the AFM order in non-superconducting materials and hope to gain further insights into the role of nematic fluctuations.

The materials studied in this work are Mn, Cr, V and Cu doped Ba-122. The high-temperature nematic susceptibilities can be all fitted by the Curie-Weiss-like function as described before \cite{LiuZ16, GuY17}, where a nematic Curie constant $A_n$ can be obtained. The value of $A_n$ is quickly suppressed in Mn, Cr and V doped samples. In Ba(Fe$_{1-x}$Cu$_x$)$_2$As$_2$, although the mean-field nematic transition temperature $T'$ becomes zero around 0.045, no nematic quantum critical point (QCP) exists since the nematic fluctuations at low temperatures are significantly suppressed. This is consistent with the fact that the AFM order becomes short-range above $x$ = 0.04 and persists up to $x$ = 0.08 as shown by the neutron diffraction measurements. Moreover, doping Mn, Cr and Cu into optimally doped BaFe$_2$(As$_{0.69}$P$_{0.31}$)$_2$ suppresses both superconductivity and nematic fluctuations. These results suggest the nematic and spin degrees of freedom are strongly coupled. Moreover, despite of different macroscopic effects of these dopants on the electronic and magnetic properties, the suppression of nematic fluctuations may be correlated with the disappearance of superconductivity. 

\section{experiments}
Single crystals of Ba(Fe$_{1-x}$$TM$$_x$)$_2$As$_2$ ($TM$ = Mn, Cr, V and Cu) and Ba(Fe$_{1-x}$$TM$$_{x}$)$_2$(As$_{0.69}$P$_{0.31}$)$_2$ ($TM$ = Mn and Cu) were grown by flux-method as reported elsewhere\cite{ChenY11,ZhangW19}. For the sake of simplicity, we will label them as $TM$-Ba122 and $TM$-BFAP, respectively. The doping levels of $TM$-Ba122 are actual doping levels determined from the inductively coupled plasma technique. The doping levels of $TM$-BFAP are nominal. The nematic susceptibility is obtained by measuring the resistance change under uniaxial pressure, which was carried out on a physical property measurement system (PPMS, Quantum Design). The uniaxial pressure is applied by a piezoelectric device as discussed previously \cite{LiuZ16}. The sign of the pressure is consistent with the experiments in hydrostaic pressure, i.e., positive and negative pressures correspond to compress and tensile a sample, respectively.
The samples were cut along the tetragonal (110) direction by a diamond wire saw and then glued on the uniaxial pressure device. Neutron diffraction experiments were carried out at Kunpeng triple-axis spectrometer (TAS) at Key Laboratory of Neutron Physics and Institute of
Nuclear Physics and Chemistry, Mianyang, China, SIKA TAS at Open Pool Australian Lightwater reactor (OPAL) \cite{sika}, Australia and Xingzhi TAS at China Advanced Research Reactor (CARR), Beijing, China, with the final energy E$_f$ = 5 meV. Cooled Be filters were used after the sample. The samples were aligned in the [H,H,L] scattering plane in the tetragonal notation. 

\section{results}

\subsection{$TM$-Ba122 ( $TM$ = Mn, Cr and V )}

\begin{figure}
\includegraphics[width=\columnwidth]{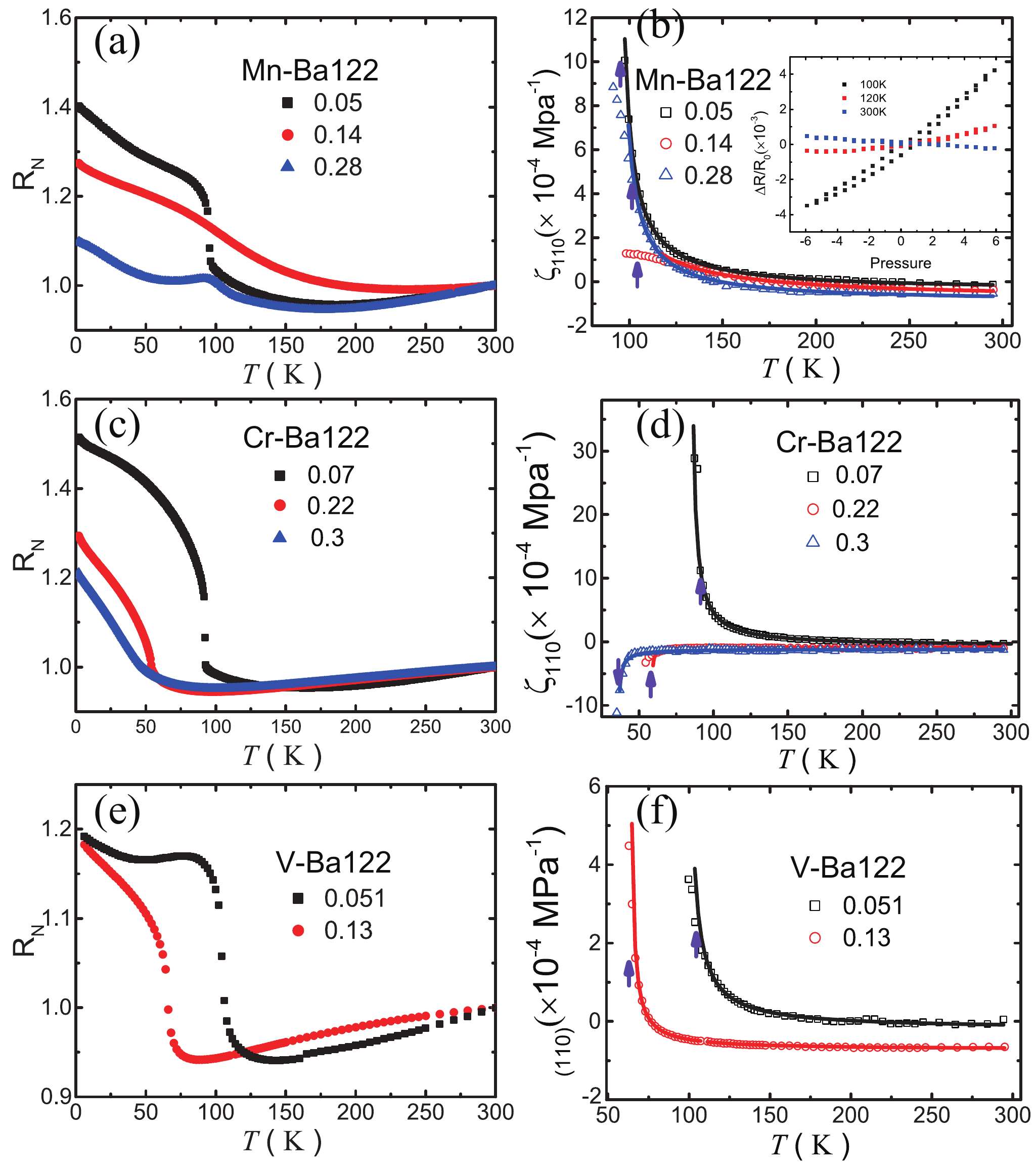}
\caption{(a) Temperature dependence of the normalized resistivity $R_N$ = $R/R_{300K}$ in Mn-Ba122. (b) Temperature dependence of $\zeta_{(110)}$ in Mn-Ba122. The inset shows the resistivity change under pressure for the $x$ = 0.05 sample at several temperatures. (c) Temperature dependence of $R_N$ in Cr-Ba122. (d) Temperature dependence of $\zeta_{(110)}$ in Cr-Ba122. (e) Temperature dependence of $R_N$ in V-Ba122. (f) Temperature dependence of $\zeta_{(110)}$ in V-Ba122. The arrows in (b), (d) and (f) indicate $T_N$ obtained from the resistivity measurements that are consistent with previous reports \cite{KimMG10,MartyK11,LiX18}. 
}
\label{fig1}
\end{figure}

We first provides results of Mn-, Cr- and V-Ba122. One of the common features of the magnetic properties in these systems is that the G-type AFM order strongly competes with the stripe AFM order \cite{KimMG10,MartyK11,TuckerGS12,InosovDS13,LiX18}. It is thus interesting to see whether their nematic fluctuations also share some similarities.

Figure 1(a) shows the temperature dependence of the normalized resistivity $R_N$ for Mn-Ba122. For the $x$ = 0.05 sample, $R_N$ exhibits a sharp upturn at $T_N$ with decreasing temperature, which is a typical behavior for samples with the substitution of Fe by 3d transition-metal elements \cite{NiN09}. It should be noted that the structural transition temperature $T_s$ is usually the same or slightly above $T_N$ in most of the samples studied here, so we will not distinguish them unless ortherwise specified. This upturn disappears in the $x$ = 0.14 sample although $T_N$ changes little, which is consistent with the suppression of the stripe AFM order \cite{KimMG10,InosovDS13}. A kink feature is found in the $x$ = 0.28 sample, and we attribute it to the AFM transition \cite{KimMG10,TuckerGS12,InosovDS13,FilsingerKA17}. 

To study the nematic susceptibility, we have measured the resistance change under the uniaxial pressure to obtain $\zeta = d(\Delta R / R _0) / dP $, where $P$ and $R_0$ are the pressure and the resistance at zero pressure, respectively, and $\Delta R = R(P)-R_0$. The subscript means that the pressure is along the tetragonal (110) direction. As discussed previously \cite{LiuZ16,GuY17}, $\zeta_{(110)}$ can be treated as the nematic susceptibility, which is analogous to the magnetic susceptibility in a paramagnetism. The inset of Fig. 1(b) gives some examples of the pressure dependence of $\Delta R / R _0$ for the $x$ = 0.05 Mn-Ba122 sample. Well above $T_N$, the resistance changes linearly with pressure, so the slope is used to calculate $\zeta_{(110)}$. When the temperature is close to $T_N$, nonlinear pressure dependence of resistance appears, which is due to the effect of large pressure \cite{MaoH18}. Since this nonlinear effect is weak, one can still roughly make a linear fit to the data. Below $T_N$, clear hysteresis behavior appears because of magnetic or nematic domains \cite{GongD17}. In this case, $\zeta_{(110)}$ is not well defined anymore and the linear fit is forced to obtain its value, which will not affect our analyses that focus on the nematic susceptibility above $T_N$.

Figure 1(b) shows the temperature dependence of $\zeta_{(110)}$ for Mn-Ba122. For the $x$ = 0.05 and 0.14 samples, $\zeta_{(110)}$ above $T_N$ can be fitted by the Curie-Weiss-like function, $A/(T-T')+y_0$, where $A$, $T'$ and $y_0$ are temperature-independent parameters \cite{LiuZ16, GuY17}. The fittings are good until the temperature is below $T_N$, which is not just due to the ill-defined $\zeta_{(110)}$ below $T_N$ but also because the nematic susceptibility cannot go infinite with decreasing temperature and should change below the phase transition as the magnetic susceptibility in a typical magnetically ordered system. For the $x$ = 0.28 sample, where the magnetic structure would have changed, $\zeta_{(110)}$ starts deviating from the Curie-Weiss-like behavior well above $T_N$.

Figure 1(c) shows the temperature dependence of $R_N$ for Cr-Ba122. Sharp upturns are found in the $x$ = 0.07 and 0.22 samples where the AFM order is still stripe-type. For the $x$ = 0.3 sample which is near the crossover from the stripe to G-type AFM order, the upturn of $R_N$ below $T_N$ becomes smooth. Figure 1(d) shows the temperature dependence of $\zeta_{(110)}$ in these samples. The nematic susceptibility in the $x$ = 0.07 behaves similar to that in the $x$ = 0.05 and 0.14 Mn-Ba122 samples. $\zeta_{(110)}$ becomes negative in the $x$ = 0.22 and 0.3 samples but can be still fitted by the Curie-Weiss-like function with negative value of $A$. This sign change has also been found in the hole-doped Ba$_{1-x}$K$_x$Fe$_2$As$_2$ system \cite{GuY17}.

Figure 1(e) and 1(f) show the results of V-Ba122. The major features are similar to those observed in Mn- and Cr-Ba122, including the sharp upturn of $R_N$ associated with $T_N$ and the Curie-Weiss-like behavior of $\zeta_{(110)}$ above $T_N$. We will further analysis the nematic susceptibility in these materials later and here we would like to point it out that there is an intimate relationship between the magnetic and nematic systems.

\subsection{Ba(Fe$_{1-x}$Cu$_{x}$)$_2$As$_2$}

\begin{figure}
\includegraphics[width=\columnwidth]{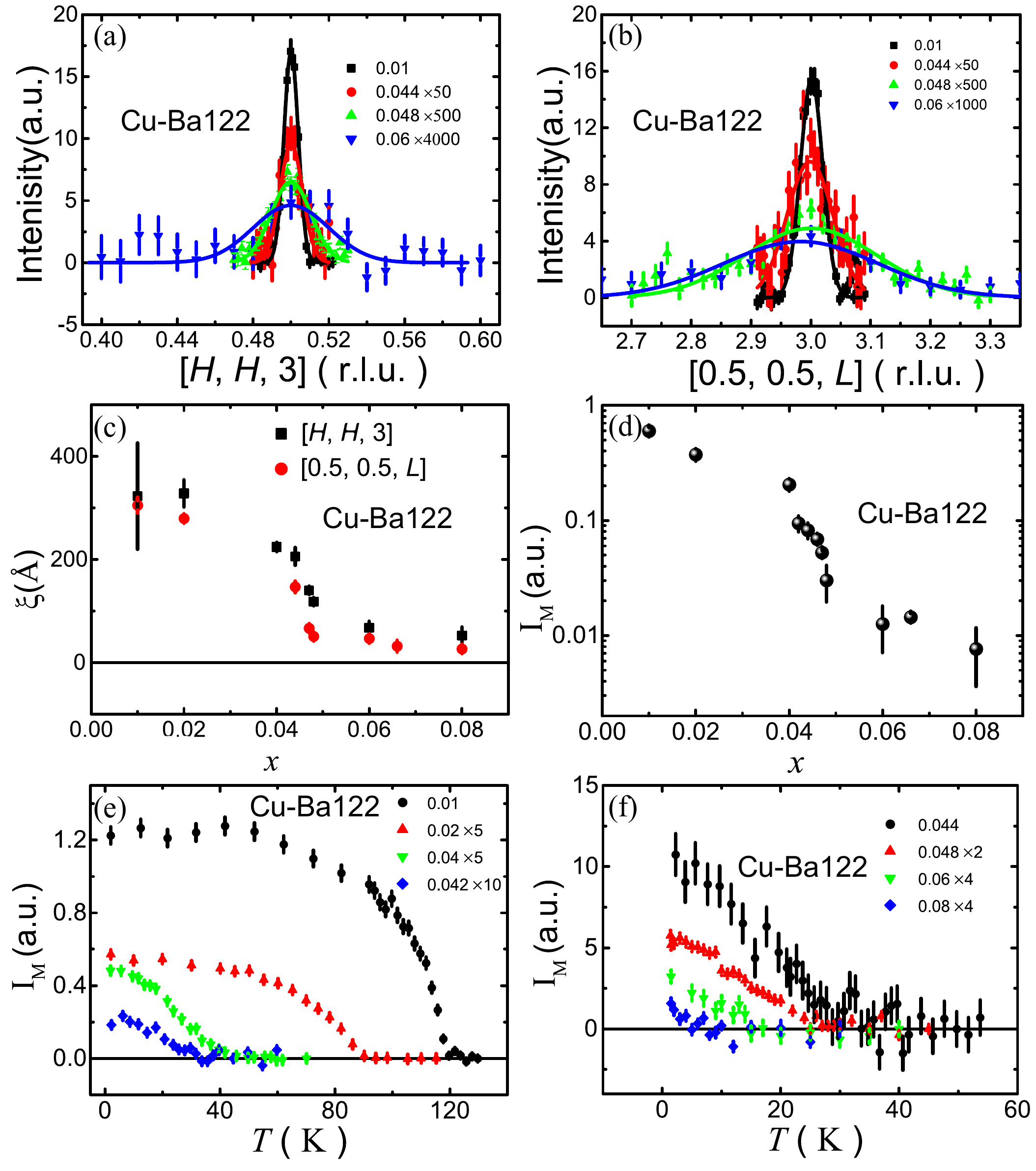}
\caption{(a) Q-scans along the [H,H,3] direction for Ba(Fe$_{1-x}$Cu$_x$)$_2$As$_2$ at 2 K. (b) Q-scans along the [0.5,0.5,L] direction for Ba(Fe$_{1-x}$Cu$_x$)$_2$As$_2$ at 2 K. The solid lines in (a) and (b) are fitted by the Guassian function. (c) Doping dependence of the magnetic correlation length along HH and L directions. (d) Doping dependence of the peak intensity at (0.5, 0.5, 3). (e) \& (f) Temperature dependence of the peak intensity at (0.5, 0.5, 3). The labels in (a), (b), (e) and (f) represent the doping levels and the scale factors of the intensities. } 
\label{fig2}
\end{figure}

The Cu doping is unique in the Ba-122 system in that the AFM order is continuously suppressed but no superconductivity is found \cite{KimMG12,TakedaH14}. Moreover, it seems that the AFM and structural transitions are separated and may disappear at different doping levels. In other words, we may expect to observe the magnetic and nematic QCPs individually without the presence of superconductivity. However, there are some discrepancies in phase diagram \cite{KimMG12,TakedaH14}. Therefore, we provide our elastic neutron scattering studies on the AFM order here, which will help us to understand the nematic fluctuations in this system. 

Figure 2(a) and 2(b) shows the HH-scans and L-scans at the magnetic peak (0.5, 0.5, 3) at 2 K for Cu-Ba122. For $x \leq $ 0.04, the magnetic peak is very sharp and the width is essentially resolution-limited. With further increasing $x$, the peak widths become much broader. This is because the AFM order becomes short-ranged as reported previously \cite{KimMG12}. To quantitatively describe the change from long-range to short-range AFM order, the doping dependence of the magnetic correlation lengths is shown in Fig. 2(c). Here the correlation $\xi$ is calculated as $\xi$ = $2\pi/FWHM$, where FWHM is the full width at half maximum obtained by the Guassian fit to the peak. For $x \leq$ 0.04, the correlation lengths along both directions are larger than 200 \AA. Since the resolution effect has not been considered, these large values of $\xi$ suggest that the AFM order is still long-range. Above $x$ = 0.04, $\xi$ quickly drops in both directions, indicating the magnetic system becomes short-range. The change from long-range to short-range AFM order is also evidenced by the doping dependence of the peak intensity as shown in Fig. 2(d), which also drastically decreases above $x$ = 0.04. Figure 2(e) and 2(f) shows the temperature dependence of the magnetic peak intensity at (0.5, 0.5, 3), which are used to determine $T_N$. We will compare the results of $T_N$ with the mean-field nematic transition temperature $T'$ later.

\begin{figure}
\includegraphics[width=\columnwidth]{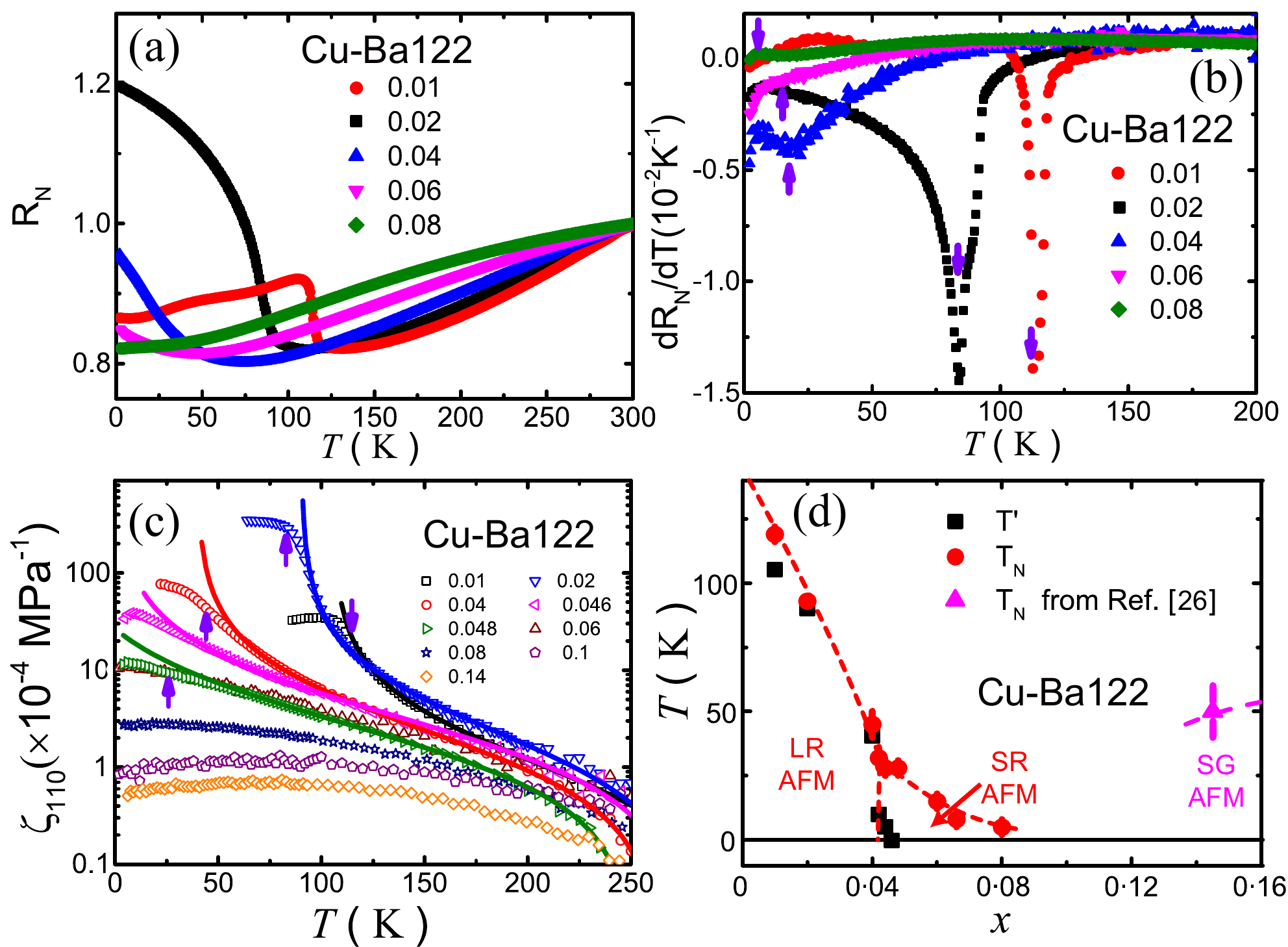}
\caption{(a) Temperature dependence of $R_N$ for Cu-Ba122. (b) Temperature dependence of $dR_N/dT$ for Cu-Ba122. (c) Temperature dependence of $\zeta_{(110)}$ for Cu-Ba122. The solid lines are the fitted results by the Curie-Weiss-like function. The arrows in (b) and (c) indicate $T_N$. (d) Doping dependence of $T_N$ and $T'$. The labels LR, SR and SG represent long-range, short-range and spin-glass, respectively. The dashed lines are guides to the eye.}
\label{fig3}
\end{figure}

Figure 3(a) shows the temperature dependence of $R_N$ for Cu-Ba122, where clear sharp upturns can be found for $x <$ 0.04. This is consistent with the above observation that the AFM order is long-range in these samples. Above $x$ = 0.04, the upturn in $R_N$ becomes smooth, which makes it hard to judge whether there is an AFM transition or not. The temperature dependence of $dR_N/dT$ shown in Fig. 3(b) also makes it clear that only the long-range AFM order can result in a dip at $T_N$. Since we have already obtained $T_N$ for $x >$ 0.04 from the neutron diffraction experiment, it seems that the short-range AFM order can still give rise to a kink feature in $dR_N/dT$.

Figure 3(c) shows the temperature dependence of $\zeta_{(110)}$ for Cu-Ba122. For $x \leq$ 0.048, the high-temperature data can be fitted by the Curie-Weiss-like function. With further increasing doping, $\zeta_{(110)}$ shows a broad hump, similar to that in overdoped BaFe$_{2-x}$Ni$_x$As$_2$ \cite{LiuZ16}. From the fittings, we can obtain the mean-field nematic transition temperature $T'$, whose doping dependence is plotted together with $T_N$ in Fig. 3(d). The short-range AFM order survives up to 0.08. At doping level higher than 0.14, the system changes into a spin-glass-like state as reported previously \cite{WangW17}. For the nematic order, $T'$ continuously decreases with increasing $x$ and becomes zero at about 0.046. 

\subsection{$TM$-BFAP ( $TM$ = Mn, Cr and Cu ) and summarized phase diagrams}

\begin{figure}
\includegraphics[width=\columnwidth]{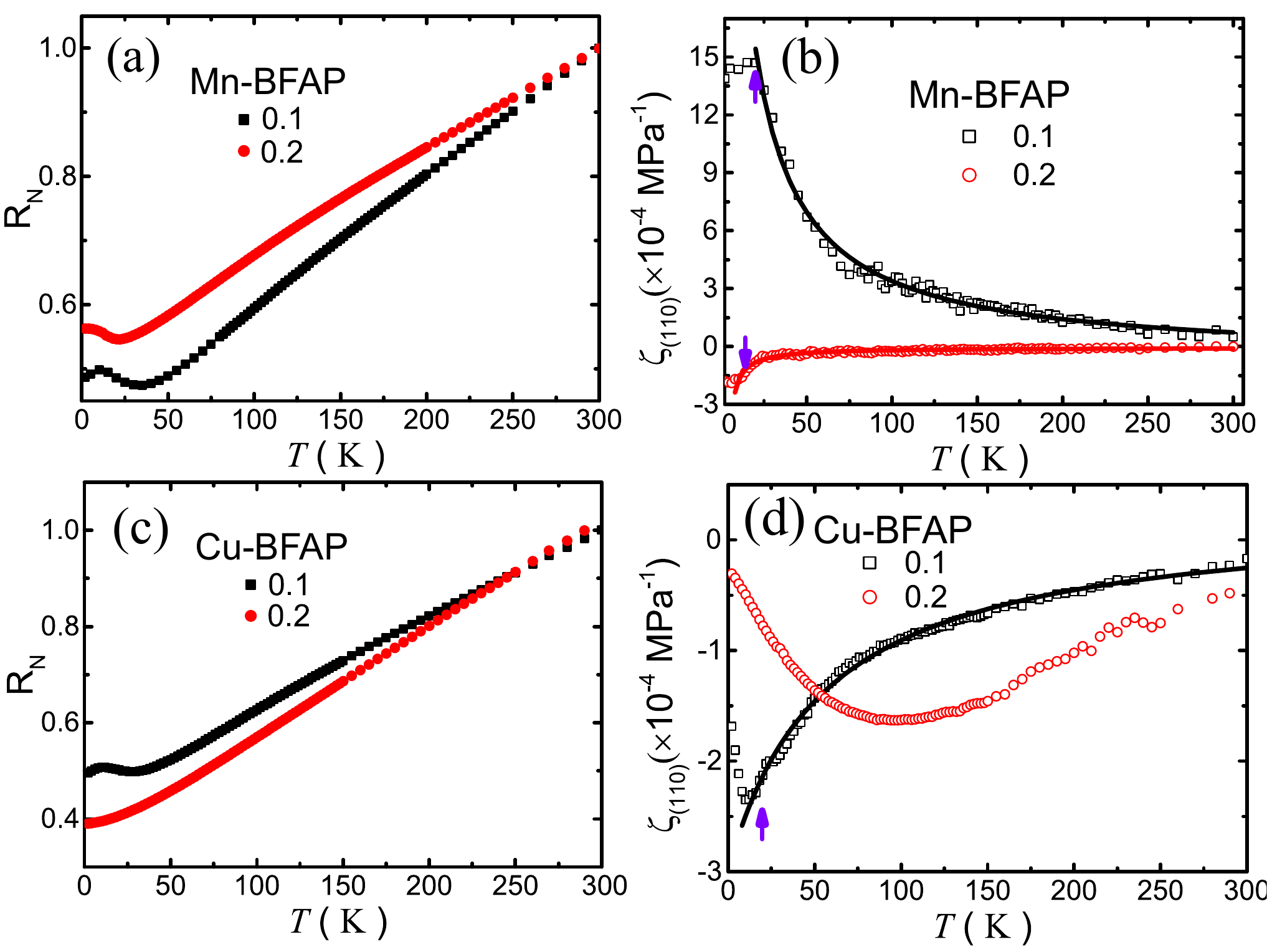}
\caption{(a) Temperature dependence of $R_N$ for Mn-BFAP. (b) Temperature dependence of $\zeta_{(110)}$ for Mn-BFAP. (c) Temperature dependence of $R_N$ for Cu-BFAP. (b) Temperature dependence of $\zeta_{(110)}$ for Cu-BFAP. The arrows in (b) and (d) indicate assumed $T_N$ from the resistivity.}
\label{fig4}
\end{figure}

In previous subsections, the effects of nonsuperconducting dopants on the nematic susceptibility in BaFe$_2$As$_2$ have been shown. In this subsection, we further show how the nematic susceptibility in optimally doped BaFe$_2$(As$_{0.69}$P$_{0.31}$)$_2$ (BFAP) changes with these dopants when the superconductivity is completely suppressed. The BFAP is chosen here as a starting material because $P$ and the $TM$ elements sit at different sites in the unit cell and their effects on superconductivity, magnetism and nematicity may be relatively easier to be separated. 

While the AFM order is nearly completely suppressed in BaFe$_2$(As$_{0.69}$P$_{0.31}$)$_2$ \cite{HuD15}, upturns appear in the temperature dependence of $R_N$ for nonsuperconducting Mn-BFAP, as shown in Fig. 4(a). The uptrun temperature decreases with increasing Mn doping level. As shown in Cr-BFAP \cite{ZhangW19}, the phosphorus doping level where the AFM order disappears changes with Cr doping. Therefore, it is assumed that the upturns are associated with the AFM transitions. This assumption is also evidenced by the temperature dependence of $\zeta_{(110)}$, which shows kinks at the same temperatures as shown in Fig. 4(b). In both the $x$ = 0.1 and 0.2 samples, $\zeta_{(110)}$ above $T_N$ can be well fitted by the Curie-Weiss-like function, but the signs are opposite. 

Figure 4(c) and 4(d) shows the temperature dependence of $R_N$ and $\zeta_{(110)}$ for Cu-BFAP, respectively. An upturn appears in $R_N$ for the $x$ = 0.1 sample, suggesting the presence of the AFM order, and disappears for the $x$ = 0.2 sample. Correspondingly, $\zeta_{(110)}$ in the $x$ = 0.1 sample can be fitted by the Curie-Weiss-like function above $T_N$ and shows a kink at it, while that in the $x$ = 0.2 sample shows a broad hump. Both signs of $\zeta_{(110)}$ are negative. For Cr-BFAP ($x$ = 0.03), the nematic susceptibility has already been studied previously \cite{ZhangW19}.

Figure 5(a) shows the doping dependence of $|A_n|^{-1}$, where $A_n$ = $\kappa A$ with $A$ as the nematic Curie constant from the Curie-Weiss-like fit of $\zeta_{(110)}$. The $\kappa$ is a phenomenological parameter associated with the effect of Fermi velocities \cite{ZhangW19}. For the system studied here, we have assumed that $\kappa$ = 1. In other words, the changes of Fermi velocities with doping are supposed to be small. This assumption will not affect the main results here. As discussed previously, the magnitude of $|A_n|$ corresponds to the strength of nematic fluctuations \cite{ZhangW19}. It has been shown in a previous study that $|A_n|^{-1}$ decreases with increasing doping level in Ni-Ba122 \cite{LiuZ16}. Similar behavior is found in Cu-Ba122. For Mn-, V- and Cr-Ba122, $|A_n|^{-1}$ increases with increasing doping level. In other words, the value of $|A_n|$ becomes much smaller than that in BaFe$_2$As$_2$, suggesting the suppression of nematic fluctuations. While the effect of Cu on $|A_n|$ is different from those of other nonsuperconducting dopants, their effects are similar in BFAP, i.e., the value of $|A_n|^{-1}$ increases significantly with all kinds of non-superconducting dopants. 

Figure 5(b) plot the relationship between the AFM ordered moment $M$ and $|A_n|^{-1}$ of these nonsuperconducting systems together with the other systems that have been reported previously \cite{GuY17}. The data for Cu-Ba122 still falls onto the same linear relationship as shown by the dashed brown line. For Mn-, Cr- and V-Ba122, a new relationship is shown by the red dashed line in Fig. 5(b), which shows that $|A_n|^{-1}$ increases with decreasing $M$. Finally, starting from the optimally doped BFAP, $|A_n|^{-1}$ also increases with Mn, Cr and Cu doping as shown by the blue dashed line in Fig. 5(b).

\section{discussions}

\begin{figure}
\includegraphics[scale=0.45]{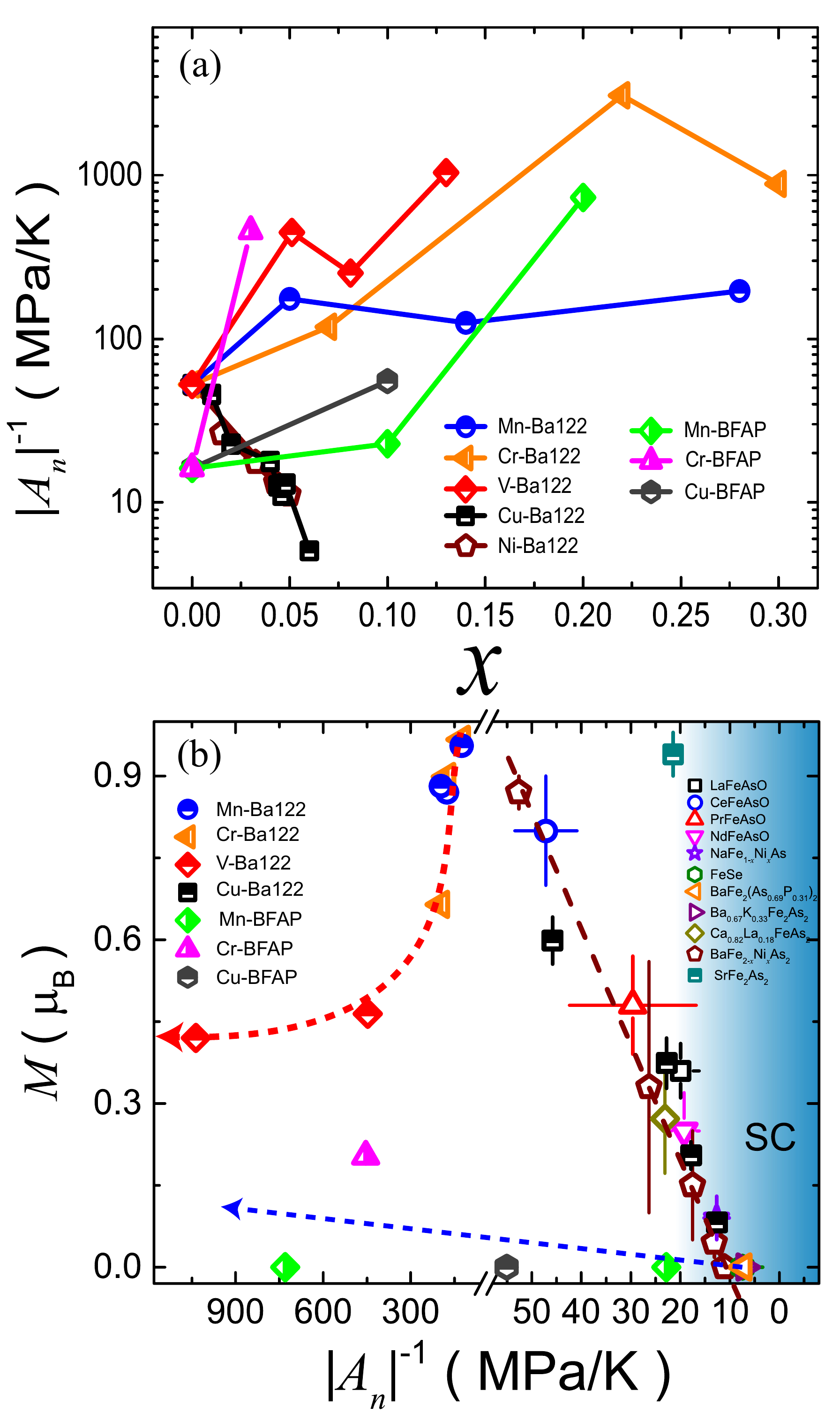}
\caption{(a) Doping dependence of $|A_n|^{-1}$ for $TM$-Ba122 and $TM$-BFAP ( TM = Mn, Cr, V and Cu ). The data for Ba(Fe$_{1-x}$Ni$_{x}$)$_2$As$_2$ and Cr-BFAP are from Ref. \cite{LiuZ16} and \cite{ZhangW19}, respectively. (b) The relationship between the AFM ordered moment $M$ and $|A_n|^{-1}$. The compounds listed in the right and the dashed straight brown line are from \cite{GuY17}. The values of $M$ for Cu-Ba122 are determined by comparing the intensities between the magnetic and nuclear peaks \cite{LuoH12}. The values of $M$ for Mn, Cr and V doped Ba-122 are estimated from Ref. \cite{KimMG10}, \cite{MartyK11} and \cite{LiX18}, respectively. Both $|A_n|^{-1}$ and $M$ of Cr-BFAP are from Ref. \cite{ZhangW19}. Since no measurements on the values of $M$ in Mn and Cr doped BFAP have been done yet, they are set as zero. The red and blue dashed lines are guides to the eye. The arrows in the red and blue dashed lines indicate increasing doping in $TM$-Ba122 ( Mn, Cr and V ) and $TM$-BFAP ( Mn, Cr and Cu ), respectively.
}
\label{fig5}
\end{figure}

We start the discussions from Mn, Cr and V dopants. In both Mn- and Cr-Ba122, the suppression of the stripe AFM order is caused by the competition from the G-type AFM order \cite{KimMG10,MartyK11,TuckerGS12,InosovDS13,YogeshS09,FilsingerKA17}. This kind of competition is also shown to exist in V-Ba122, where a spin-glass state appears first after the stripe AFM order is completely suppressed above $x$ = 0.25 \cite{LiX18}. Our results in Fig. 5(a) show that Mn, Cr and V also have the similar effects on the nematic susceptibility, i.e., they all result in the increase of $|A_n|^{-1}$. Moreover, the rate of this increase is largest for Cr doping, which accords with that the suppression of $T_N$ of the stripe AFM order is also fastest in Cr-Ba122. Since the G-type AFM order does not break the $C_4$ rotational symmetry, it is reasonable to conclude that the suppression of nematic fluctuations in these systems is due to the competition from the G-type AFM spin fluctuations. It is thus not surprising that in both Mn- and Cr-BFAP, $|A_n|^{-1}$ also increases significantly with increasing doping levels as shown in Fig. 5(a).

The Cu doping has different effects on the magnetic system in BaFe$_2$As$_2$ from the above non-superconducting dopants. First, the stripe AFM order is suppressed but is not replaced by the G-type order. In heavily overdoped Cu-Ba122 ( $x \geq$ 0.145 ), the magnetic system is spin-glass-like and the AFM order is short-range and stripe-type \cite{WangW17}. At first glance, this is the same as what we observed here for the samples with 0.044 $\leq x \leq$ 0.08. However, $T_N$ decreases and increases with increasing doping for $x \leq$ 0.08 and $x \geq$ 0.145, respectively. The difference of magnetic orders in these two doping regions can also be found from the doping dependence of the integrated magnetic intensity, which increases with doping for $x \geq$ 0.145 but decreases for $x \leq$ 0.08. It has been suggested that the certain arrangements of Fe and Cu may favor a magnetically ordered state \cite{WangW17}, as also seen in heavily overdoped NaFe$_{1-x}$Cu$_x$As \cite{SongY16}. In our case, the content of Cu may be too little to introduce any kind of particular Fe-Cu arrangement. Therefore, the short-range AFM order below $x$ = 0.08 should be the result of randomly distributed Cu dopants, which may locally disturb the long-range stripe AFM order and make it short-range. Another possibility is that the electron doping from Cu dopants may result in short-range AFM order. We note that the short-range AFM order has also been found in the electron-doped systems of Co- and Ni-Ba122 \cite{LuoH12,LuX14,PrattDK11}, although it is incommensurate. 

The effects of Cu doping on nematic fluctuations in Cu-Ba122 can also be understood by the local effects of Cu dopants. At one hand, the Cu doping results in the increase of $A_n$ ( Fig. 5(a) ), suggesting the enhancement of nematic fluctuations. On the other hand, the nematic susceptibility does not follow the Curie-Weiss-like behavior for samples with $T'$ close to zero at low temperatures ( Fig. 3(c) ), which suggests that the nematic fluctuations are significantly suppressed by Cu dopants. This seemingly contradictory results may be explained as two different effects of Cu dopants at different temperature regions. First, the nematic fluctuations will be enhanced if the magnetic order is suppressed with the stripe-type AF fluctuations maintained, as shown by the very similar doping dependence of $A_n$ and $T'$ in Cu- and Ni-Ba122 \cite{LiuZ16,GuY17}. This effect may be treated as a global effect of dopants to the nematic system. Second, the strong local effects of Cu dopants may prevent the nematic susceptibility from further increasing at low temperatures. Since the nematic order is directly coupled to the lattice, it is not surprising that $\zeta_{(110)}$ cannot be infinite when $T'$ is close to zero because it would mean that the lattice will become unstable. Naively, it may suggest that Cu dopants may limit the nematic correlations and keep the nematic system from long-range ordering. Further high-resolution structural studies may give an answer to this hypothesis. It is interesting to note that many superconducting systems may have an avoided nematic QCP near optimally doping levels where $T'$ becomes zero \cite{LiuZ16,KuoHH16,GuY17}. Similarly, an avoided nematic QCP may also present in Cu-Ba122, i.e., quantum nematic fluctuations could dominate at high temperatures when $T'$ = 0 but there is no actual nematic QCP.

Our results suggest a close relationship between the stripe AFM order and the nematicity. It has been shown that the ordered moment has a roughly linear relationship with $|A_n|^{-1}$ for many parent compounds and doped superconducting samples, as shown in Fig. 5(b). It suggests that the nematic fluctuations are enhanced when the stripe AFM order is suppressed. For Cu-Ba122 which does not induce any other type of magnetic order competing with the stripe AFM order, this relationship still holds. However, for Mn-, Cr- and V-Ba122, $|A_n|^{-1}$ increases drastically with the suppression of the stripe AFM order as shown by the red dashed line in Fig. 5(b), which clearly suggests that the competition from G-type AFM order suppresses the nematic fluctuations. This results, together with the linear relationship between $|A_n|^{-1}$ and $M$ in other systems, indicate that the stripe AFM order is directly associated with nematic fluctuations. 

Our results also suggest a close relationship between the nematic fluctuations and superconductivity. We have suggested in a previous work that superconductivity appears only when $|A_n|^{-1}$ is small enough \cite{GuY17}. This is consistent with the results in this work that $|A_n|^{-1}$ is drastically increased with non-superconducting dopants doping into the superconducting samples. Moreover, whether a dopant can lead to superconductivity in BaFe$_2$As$_2$ seems associated with whether it can enhance nematic fluctuations. A particular case is Cu-doped Ba-122, where Cu doping only suppress low-temperature nematic fluctuations. Of course, it is not to say that the superconductivity must come from nematic fluctuations since one mechanism may affect both superconductivity and nematicity simultaneously. Still, our results put nematicity as one of the key aspects to understand superconductivity in iron-based superconductors. 

\section{conclusions}
We have systematically studied the doping effects of Mn, Cr, V and Cu on the nematic susceptibility in parent compound Ba-122 and optimally-doped superconducting BFAP. The nematic Curie constant $|A_n|$ are drastically decreased in all cases except in Cu-Ba122. In the latter, nematic fluctuations are only suppressed at low temperatures. Combining our previous studies, our results suggest that the stripe AFM order and superconductivity may both have intimate relationship with nematic fluctuations. In other words, nematicity may play one of the key roles in the low-energy physics of iron-based superconductors.

\begin{acknowledgments}
This work was supported by the National Key R\&D Program of China (Grants No. 2017YFA0302900 and No. 2016YFA0300502,2016YFA0300604), the National Natural Science Foundation of China (Grants No. 11874401 and No. 11674406, No. 11374346, No. 11774399, No. 11474330, No. 11421092, No. 11574359 and No. 11674370), the Strategic Priority Research Program(B) of the Chinese Academy of Sciences (Grants No. XDB25000000 and No. XDB07020000, No. XDB28000000), China Academy of Engineering Physics (Grant No. 2015AB03) and the National Thousand-Young Talents Program of China. H. L. is grateful for the support from the Youth Innovation Promotion Association of CAS (2016004).
\end{acknowledgments}

\end{document}